\documentclass[twoside,twocolumn,9pt]{article}
\usepackage{extsizes}
\usepackage[super,sort&compress,comma]{natbib} 
\usepackage[version=3]{mhchem}
\usepackage[left=1.5cm, right=1.5cm, top=1.785cm, bottom=2.0cm]{geometry}
\usepackage{balance}
\usepackage{times,mathptmx}
\usepackage{sectsty}
\usepackage{graphicx} 
\usepackage{lastpage}
\usepackage[format=plain,justification=justified,singlelinecheck=false,font={stretch=1.125,small,sf},labelfont=bf,labelsep=space]{caption}
\usepackage{float}
\usepackage{fancyhdr}
\usepackage{fnpos}
\usepackage[english]{babel}
\usepackage{array}
\usepackage{droidsans}
\usepackage{charter}
\usepackage[T1]{fontenc}
\usepackage[usenames,dvipsnames]{xcolor}
\usepackage{setspace}
\usepackage[compact]{titlesec}
\usepackage{hyperref}
\usepackage{amsmath} 
\usepackage{gensymb}
\usepackage{color}
\usepackage{lipsum}
\usepackage{amssymb}

\usepackage{epstopdf}

\definecolor{cream}{RGB}{222,217,201}

\begin{document}


\makeFNbottom
\makeatletter
\renewcommand\LARGE{\@setfontsize\LARGE{15pt}{17}}
\renewcommand\Large{\@setfontsize\Large{12pt}{14}}
\renewcommand\large{\@setfontsize\large{10pt}{12}}
\renewcommand\footnotesize{\@setfontsize\footnotesize{7pt}{10}}
\makeatother

\renewcommand{\thefootnote}{\fnsymbol{footnote}}
\renewcommand\footnoterule{\vspace*{1pt}%
\color{cream}\hrule width 3.5in height 0.4pt \color{black}\vspace*{5pt}} 
\setcounter{secnumdepth}{5}

\makeatletter 
\renewcommand\@biblabel[1]{#1}            
\renewcommand\@makefntext[1]%
{\noindent\makebox[0pt][r]{\@thefnmark\,}#1}
\makeatother 
\renewcommand{\figurename}{\small{Fig.}~}
\sectionfont{\sffamily\Large}
\subsectionfont{\normalsize}
\subsubsectionfont{\bf}
\setstretch{1.125} 
\setlength{\skip\footins}{0.8cm}
\setlength{\footnotesep}{0.25cm}
\setlength{\jot}{10pt}
\titlespacing*{\section}{0pt}{4pt}{4pt}
\titlespacing*{\subsection}{0pt}{15pt}{1pt}

\linespread{1.5}
\large

\makeatother

\twocolumn[
  \begin{@twocolumnfalse}
\vspace{3cm}
\sffamily
\begin{tabular}{m{1.5cm} p{15.5cm} }

 & \noindent\LARGE{\textbf{\textit{In-Situ} Growth of Graphene on Hexagonal Boron Nitride for Electronic Transport Applications}} \\
\vspace{0.3cm} & \vspace{0.3cm} \\

 & \noindent\large{Hadi Arjmandi-Tash$^{\ast}$}\\

& \noindent\normalsize{Transferring graphene flakes onto hexagonal boron nitride (h-BN) has turned as a common approach for fabricating graphne/h-BN heterostructures. Controlling the alignment between graphene and h-BN lattices can hardly be achieved and the h-BN/graphene interface is prone to be contaminated in this elaborate process. Direct synthesis of graphene on h-BN is a rapidly growing alternative. \textit{In situ} grown graphene is individually tailored to conform the specific h-BN flake, hence the limitations of the conventional fabrication approach are overcome. Developed processes pave the initial steps to improve the scalablity of the device fabrication for industrial applications. Reviewing the developments in the field, from the birth point to the current status is the focus of this letter. I show how the field is being progressed to overcome the existing challenges one after the other and discuss where the field is heading to.}  

\end{tabular}

\end{@twocolumnfalse} \vspace{0.6cm}

]

\renewcommand*\rmdefault{bch}\normalfont\upshape
\rmfamily
\section*{Introduction}


\footnotetext{\textit{$^{\ast}$~Kavli Institute of Nanoscience, Delft University of Technology, Lorentzweg 1, 2628 CJ Delft, The Netherlands. E-mail: h.arjmanditash@tudelft.nl}}





Electronic mobility in supported graphene is highly limited by the roughness and the charged impurities on the surface of the substrate\,\cite{Hwang2007, Du2008, Mayorov2012}. Inserting a thick ($\gtrsim 10\,nm$) buffer layer of hexagonal boron nitride -- with atomically flat and neutral surface -- in between graphene and the supporting substrate helps to overcome those limitations. Heterostructure of graphene on multilayer h-BN flakes was first successfully realized and studied by C. Dean et al\,\cite{Dean2010}; their device exhibited charge carrier mobilities as high as 140,000 $cm^2/V\,s$ which was not reachable on supported devices by that time. Next work from the same group\,\cite{Petrone2012} also confirmed the advantages of h-BN as a matress for CVD graphene. Lately by sandwiching graphene between two h-BN layers, long mean free paths (comparable to the device size) and ballistic transport was reported even at room temperature\,\cite{Mayorov2011}\cite{Wang2013a}. Its dielectric nature and lattice parameter -- which is close to that of graphene -- are the other remarkable properties of h-BN as a matress for graphene. 

In all these reports, chemical vapor deposition (CVD) or exfoliated graphene is first isolated on an intermediate substrate and then transferred onto the h-BN flakes. The process is of remarkable disadvantages: \textit{i)} contaminating the surface of h-BN and graphene is highly possible during this process as air or water molecules might be trapped at the interface. \textit{ii)} graphene can be damaged or wrinkled and \textit{iii}) the process does not provide any control over the relative orientation of graphene and h-BN: the random orientation of the lattices may lead to irreproducible results. \textit{iv}) Macroscopic alignment of microscale graphene on h-BN flakes is another issue which is time-consuming and troublesome in practice. 

Transfer-free, direct growth of graphene on h-BN techniques offer solutions to overcome the limitations listed above. Particularly, in such techniques  the graphene/h-BN interface is realized \emph{in-situ} and thus no external contaminant may get trapped in between.
  
In this letter, the progress of the techniques to grow graphene on thick h-BN flakes is reviewed. Standard CVD of graphene benefits from the presence of a catalyst (e.g. copper) as the promoter of the growth; the absence of such a catalyst hinders direct growing of graphene on h-BN. I discuss how different approaches tackled this limitation. Different aspects of such growth methods are reviewed here. {\color{black}Note that there are some methods developed for growing both graphene and mono- (few-) layer h-BN together to make thin ($\lesssim 10\,nm$) graphene/h-BN stacks or patchworks\,\cite{Wang2013,Roth2013,Liu2011a,Kim2013b, Entani2019, Wu2018, Song2016}; Interestingly, the growth of the first sample of this type was reported even before the first isolation of graphene by exfoliation\,\cite{Oshima2000}. Thickness of h-BN layers achieved in such approaches, however, are not enough to smoothen the roughness of underlying substrates and diminish the effect of the random potentials resting on the wafer; such approached are out of the scope of this letter then.}

\section{Growth Yield}
Chemical growth of graphene relies based on the decomposing carbon-rich precursor molecules (\textit{e.g.} methane) at the presence of a catalyst. The elevated temperature of the growth chamber provides enough energy for the reouultant carbon atoms to get mobilized and reach and bind the other carbon atoms and form a graphene layer. Indeed the presence of the catalyst plays a vital role to speed-up the growth: skipping the catalyst is a major limitation for the growth of graphene on arbitrary (\textit{i.e.} non-catalyst) substrates such as on h-BN. 

The first paper about directly growing CVD graphene on h-BN was submitted for publication just four months after the first realization of the graphene/h-BN stacking\,\cite{Dean2010}. The rapid inception implies the importance of the \textit{in-situ} growth approaches in the first place. Published by Ding \textit{et al}\,\cite{Ding2011}, this paper confirmed the possibility of chemically growing few-layer graphene on h-BN powders. The importance of catalyst was, however, overlooked: no remarkable arrangement was considered to compensate its absence. The size of the graphene domains remained unclear; even-though as a normal CVD process with short growth time (compared to the later reports) was used, domains larger than few nanometers are hardly expected. \autoref{fig: DG_literature_review_1}-a and b shows some of the results. Lately and as the field started developing, few approaches have been introduced to overcome the absence of the catalyst. 

\subsection{Elongated Growth}
\begin{figure}[t]
\centering
\includegraphics[width=0.4\textwidth]{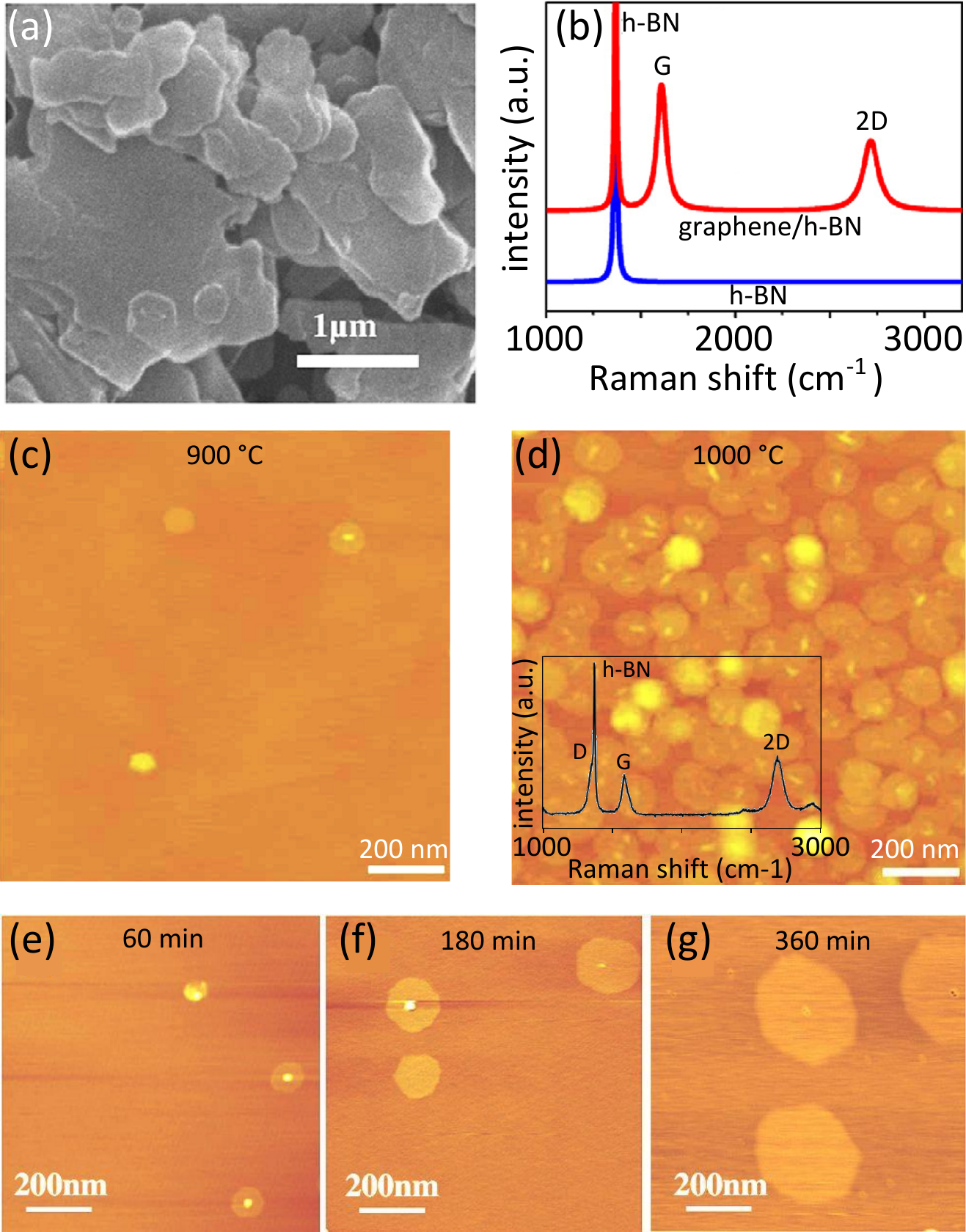} 
\centering
\caption{Early reports on the direct growth of graphene on thick h-BN flakes\\
a) and (b) Results of Ding \textit{et al}: Graphene was grown on h-BN powders through a CVD process. The Raman spectra on (b) are taken from bare h-BN powder and after the growth. Reprinted from \,\cite{Ding2011}, Copyright (2011), with permission from Elsevier.\\
c) and (d) Results of Son \textit{et al}: The effect of the growth temperature on the density of the obtained flakes was reported as AFM mappings in this work. The density raised a lot as the temperature increased from 900\degree C to 1000\degree C. This growth lasts for 2 hours. The inset in (d) shows an example of the Raman spectrum reported on this sample. Adapted from\,\cite{Son2011} with permission of The Royal Society of Chemistry.\\
e), (f) and (g) Results of Tang \textit{et al}: AFM measurements shown in these figures were performed on the samples with 1, 3 and 6 hours of growth. Reprinted from\,\cite{Tang2012}, Copyright (2012), with permission from Elsevier.   
}
\label{fig: DG_literature_review_1}
\end{figure}

Elongating the growth course is the simplest approach to compensate the absence of the catalyst. Son \textit{et al}\,\cite{Son2011} mechanically exfoliated h-BN flakes on a silicon wafer and grew graphene in an atmospheric pressure CVD chamber. Different growth temperatures ranging between 900\degree C to 1000\degree C with a similar growth duration of 2 hours were tested. Domains of $\sim 100\,nm$ in diameter achieved. They reported a direct trend of increasing the density of the graphene pads upon increasing the growth temperature (\autoref{fig: DG_literature_review_1}-c and d). The grown graphene flakes were of rounded shapes with the thicknesses of the order of 0.5\,nm. AFM, Raman and XPS analysis have been performed to confirm the growth and characterize the graphene. 

The approach was followed later by Tang \textit{et al}\,\cite{Tang2012}. Like the Son's experiment, graphene was grown on hexagonal boron nitride flakes exfoliated on silicon wafer, albeit through a low pressure CVD process. They noticed that screw dislocations on the flakes are favorable nucleation sites. The slow growth rate due to the missing of the catalyst was evident in the results reported; a growth duration of 6 hours only led to the formation of graphene grains of maximum 270\,nm in diameter (\autoref{fig: DG_literature_review_1}-e to g). The graphene domains were mostly single layer. 

Elongating the growth -- although is simple -- is of certain drawbacks: The reported graphene domains -- even after several hours of growth -- could hardly reach few hundreds of nanometers and are incompatible with typical device fabrication processes. The long operation time and high energy consumption are unfavorable for industrial applications. 

\subsection{Optimized Growth Parameters}
Fine-tuning the growth parameters is a wiser approach to improve the yield.  Mishra \textit{et al}\cite{Mishra2016} showed that in a typical CVD process, increasing the partial pressure of hydrogen and elevating the growth temperature can achieve an improved growth rate of up to $100\,nm/min$ (to be compared with $5\,nm/min$ with standard CVD growth parameters\,\cite{Tang2015})(\autoref{fig: DG_literature_review_4}-a). They systematically investigated the effect of the partial pressure of hydrogen on the crystallinity of the grown graphene domains: Low partial pressure of hydrogen ($H_2$ to $CH_4$ ratio of 1:1) mainly achieved circular poly-crystalline grains. Increasing the partial pressure of hydrogen ($H_2$ to $CH_4$ ratio of 30:1), however, led to the formation of van der Waals epitaxy (discussed in \autoref{sec: van der Waals epitaxy}) with aligned hexagonal grains. Indeed reducing the density of nucleation centers via hydrogen etching plays a vital role in reported results. 
\subsection{Plasma-Enhanced Chemical Vapor Deposition}
Inspired by earlier works on the growth of graphene on non-catalyst substrates\,\cite{Zhang2011}\cite{Zhang2012b}, Yang \textit{et al} employed remote plasma-enhanced chemical vapor deposition (RPE-CVD) technique to grow graphene on exfoliated h-BN flakes\,\cite{Yang2013a}. Here a remote plasma source decomposes methane molecules into various 
reactive radicals prier reaching the substrate; hence catalyst can be omitted. The approach provides enough control over the number of layers and uniformity of graphene. In principal, the size of the graphene is only limited by the size of underlying h-BN flakes. Growth temperature controls both the epitaxy and the rate of the growth: Even-though increasing the growth temperature improves the growth rate, population of nucleation centers also increases at the same time which may  lead to three-dimensional -- instead of layer by layer -- growth and suppress the epitaxy. The growth temperature of $\sim$500\degree C was the best compromise between epitaxy and growth rate. In this condition, several growth periods, each of two to three hours are still required to obtain the desired graphene sizes. 

\subsection{Gaseous Catalyst}
While the presence of the h-BN as a background substrate does not leave any room for a solid catalyst, Tang \textit{et al}\,\cite{Tang2015} used gaseous silane ($SiH_4$) and germane ($GeH_4$) catalysts to boost the growth. At a temperature of 1280\degree C, the growth rate reached $50\,nm/min$ and $400\,nm/min$, respectively in the presence of  germane and silane: the yield was highly improved comparing to the recorded $5\,nm/min$ in the absence of any catalyst. They noticed that elevating the growth temperature upto 1350\degree C further accelerate the growth to reach $\sim 1\,\mu m/min$ (\autoref{fig: DG_literature_review_4}-b). Importantly , the Auger electron spectroscopy of the domains did not exhibit any trace of silicon or germanium in the grown graphene crystal. Apparently, those atoms only stick to the edge of the graphene domains and lower the reaction barrier for carbon precursors to form the honeycomb lattice during the growth (\autoref{fig: DG_literature_review_4}-c). AFM analyses confirmed that more than $93\,\%$ of the graphene domains are well oriented with respect to the background lattice of h-BN.

\subsection{Proximity-Driven Overgrowth}
\begin{figure}[p]
\centering
\includegraphics[width=0.4\textwidth]{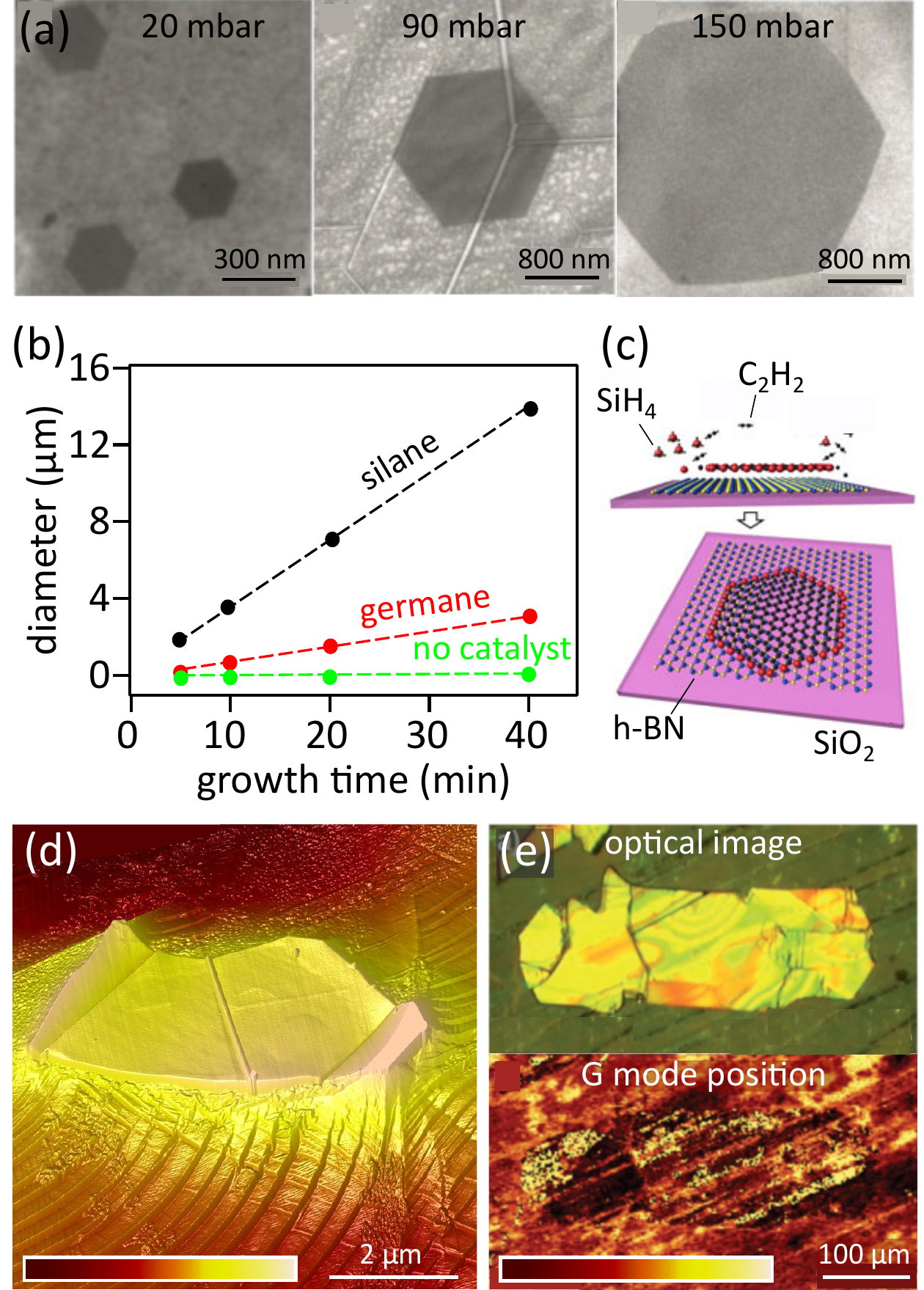} 
\centering
\caption{Various approaches to improve the growth rate\\
a) Effect of increasing the growth pressure on the grin size: SEM images showing graphene grains synthesized on h-BN flakes at 1150\degree C under various chamber pressures. Reprinted from\,\cite{Mishra2016}, Copyright (2016), with permission from Elsevier.\\
b and c) Gaseous phase catalyst to improve the growth rate: b) comparison of the experimentally measured grain size of graphene flakes grown at 1280\degree C with and without gaseous catalysts. c) Schematic illustration showing the mechanism of growing monolayer graphene onto h-BN: silicon atoms (shown as red color spheres) achieved from the decomposition of $SiH_4$ bind to the edge of the graphene and boost the growth. Reprinted with adaptations from\,\cite{Tang2015}.\\
d and e) Proximity driven over growth of graphene onto h-BN flakes pre-exfoliated on the copper foil: d) AFM mapping showing an h-BN flake (covered with graphene) on the copper foil at the end of the growth course. The color code shows the height, ranging between $0\,nm$ and $500\,nm$. e) Optical image (top) and Raman G mode position mapping (bottom) of a millimeter scale h-BN flake on the copper foil, fully covered by graphene. The color code shows the Raman frequency ranging between $1584\,cm^{-1}$ and $1600\,cm^{-1}$. Reprinted with adaptations from\,\cite{Arjmandi-Tash2017}.\\
}
\label{fig: DG_literature_review_4}
\end{figure}

A new approach in the growth of graphene on non-catalyst materials such as h-BN was introduced by our group recently\,\cite{Arjmandi-Tash2017}. Unlike the previous approaches, the growth is performed on h-BN flakes which are pre-exfoliated on the copper foil (\autoref{fig: DG_literature_review_4}-d). Hence, the carbon-rich precursors still have access to catalyst, albeit indirectly. The growth rate of graphene on h-BN was improved dramatically: a full coverage of graphene on millimeter sized h-BN flake is achievable in the same rate of graphene on surrounding copper foil (\autoref{fig: DG_literature_review_4}-e). The obtained devices exhibited charge carrier mobilities of $20,000\,cm^2/Vs$ and very neutral grapheneh-BN interfaces. 

\section{Growth Mechanism}\label{sec: van der Waals epitaxy} 
\begin{figure}[t]
\centering
\includegraphics[width=0.4\textwidth]{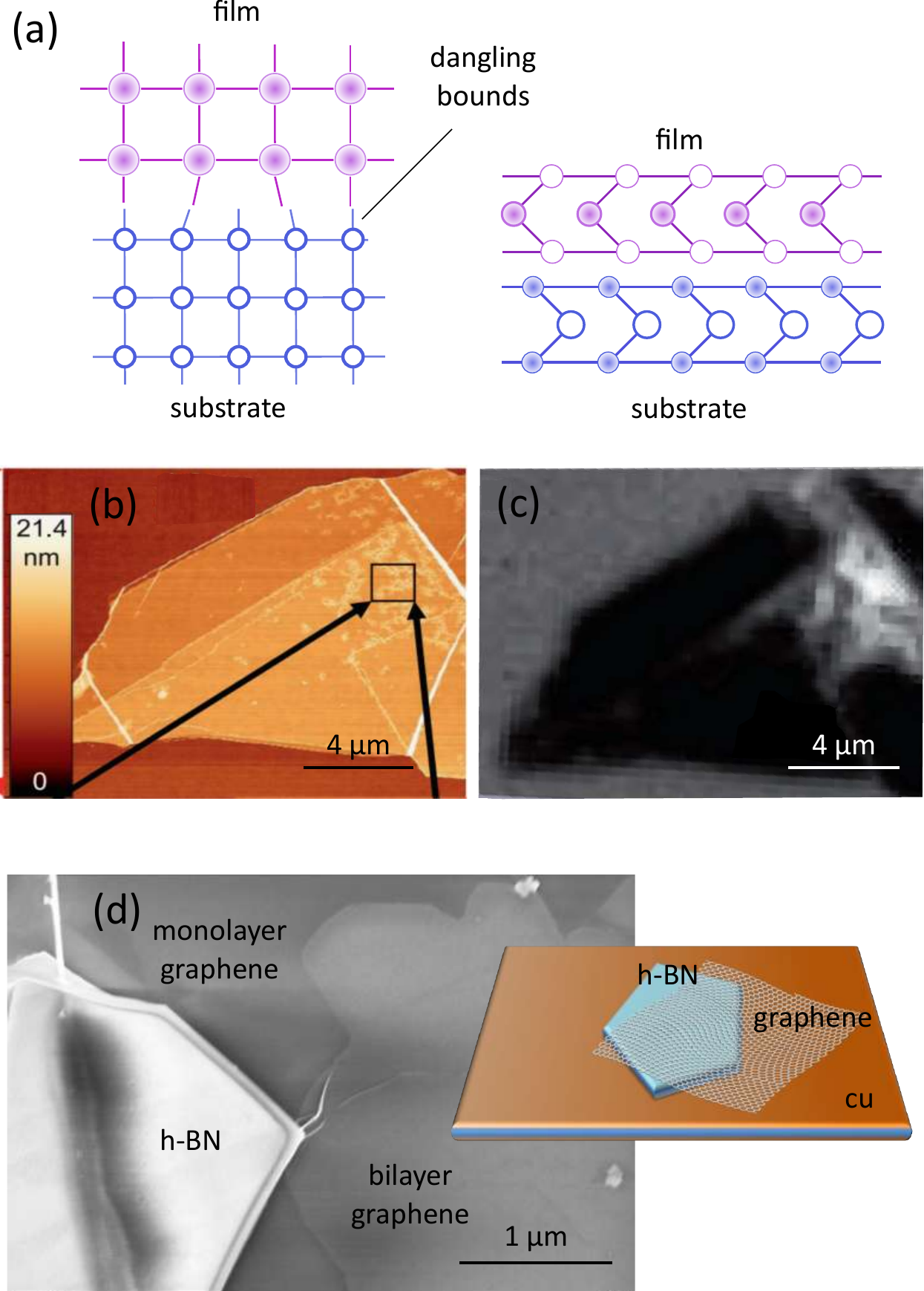} 
\centering
\caption{Known mechanisms governing the growth of graphene on h-BN flakes\\
a) Comparison of the epitaxial growth in between materials with (left) and without (right) dangling bonds, the latter has been dubbed \emph{van der Waals epitaxy}; in both cases the atoms, corresponding to the substrate and to the deposited film are shown in blue and pink respectively. \\
b) AFM and c) Raman (intensity of 2D peak of graphene) mapping of a h-BN flake with deposited graphene via molecular beam epitaxy; the mappings show that carbon atoms are freely migrated and accumulated in some preferential spots on the surface of h-BN. Reprinted from\,\cite{Garcia2012a}, Copyright (2012), with permission from Elsevier.\\
d) Schematic representation of the mechanism (inset) and SEM image of graphene (both mono- and bi-layer) nucleated on the copper foil and extended over the h-BN flake via proximity driven growth.\\
} 
\label{fig: DG_literature_review_2}
\end{figure}

In fabricating heterostructures by depositing a material on a substrate at least two mechanisms are considerable: In materials with terminating layers full of dangling bonds, atoms in one material can establish covalent bounds only to the atoms of similar lattice; indeed the covalent bonds are sensitive to the length and the angle between the atoms. This is an important hindrance in epitaxial growth between materials with different lattice parameters. \autoref{fig: DG_literature_review_2}-a schematically illustrates this mechanism. The situation is different in between two materials with perfect terminating surfaces and no dangling bonds (\autoref{fig: DG_literature_review_2}-b). Here, the absence of the dangling bonds can lead to the formation of sharp interfaces with small amount of defects. This growth mechanism is referred to as \emph{van der Waals epitaxy} and can be realized even in the presence of large lattice mismatches\,\cite{Ueno1997}\cite{Sasaki1997}. As graphene and h-BN are both free from dangling bonds, van der Waals epitaxy is normally the governing growth mechanism.

Garcia \textit{et al}\,\cite{Garcia2012a} first reported van der Waals epitaxy in \textit{in-situ} growing graphene on h-BN. Unlike the previous experiments, graphene was grown by molecular beam epitaxy (MBE) using solid carbon sources. Combined Raman and AFM analysis revealed that the growth is independent of the flux of carbon atoms; instead, carbon atoms deposed on the surface, migrate freely and accumulate in selective spots on the h-BN surface (\autoref{fig: DG_literature_review_2}-b and c). This observation pointed out that the carbon atoms are of high mobilities on the neutral h-BN, confirming that van der Waals epitaxy was achieved. Recently, van der Waals epitaxy was reported in CVD growth of graphene on h-BN also\,\cite{Mishra2016}. {\color{black} The lack of any covalent bond results in weak interfacial interaction between graphene and h-BN\,\cite{Entani2019}}   

A seprate but complementary growth mechanism involves extending graphene -- already nucleated on the copper foil -- onto nearby h-BN flakes. This mechanism was first observed in graphene grown on few-layer chemically grown h-BN sheets\,\cite{Kim2013b}; our recent work\,\cite{Arjmandi-Tash2017}, however, confirmed that the thickness of the h-BN is not any limitation as graphene can overgrow on hundreds-of-nanometers-thick h-BN flakes, mechanically pre-exfoliated on the copper foil. Inset to \autoref{fig: DG_literature_review_2}-d explains our hypothesized model for this growth. Precursors are cracked on the copper foil. The achieved carbon radicals move randomly in different directions and are energetic enough to continuously jump over the h-BN flake and bond as-growing graphene. The presence of the copper catalyst and high mobility of carbon atoms -- as a result of the van der Waals epitaxy -- guarantees high growth rate atop h-BN flakes. The main panel of \autoref{fig: DG_literature_review_2}-d shows SEM image of a h-N flake covered with graphene. 

\section{Epitaxial Growth}
Once two similar patterns of crystalline lattices (\textit{e.g.} graphene and h-BN lattices) are superimposed with a small displacement or rotation in between, a secondary pattern known as moir\' e generates\,\cite{Xue2011}\,\cite{Decker2011}. Moir\'e superlattice potential affects the propagation of charge carriers in graphene by inducing new Dirac points, known as \emph{satellite Dirac points (SDPs)} in the band structure of graphene.  The energy of SDPs ($E_{SDP}$) depends on the wave vector of the superlattice ($\lambda_{SL}$) which itself is a function of the misorientation angle ($\Phi$) between the lattices:\,\cite{Yankowitz2012}\\[8pt]
$E_{SDP}=\pm \frac{\hbar v_f|\vec{G}|}{2}=\pm \frac{2\pi \hbar v_F}{\sqrt{3}\lambda_{SL}}$, where:\\[8pt]
$\lambda_{SL}=\frac{(1+\delta)a}{\sqrt{2(1+\delta)(1-cos\Phi)+\delta^2}}$\\
In this relation, $\vec{G}$ represents the reciprocal superlattice vector and $v_F\approx 10^6\,m/s$ is the Fermi velocity of quasiparticles in graphene. Additionally, $a=2.46\,\AA$ and $\delta=1.8\%$ are the lattice parameter of graphene and the mismatch between graphene and h-BN lattices, respectively. 

Controlling and minimizing the misorientation angle in graphene/h-BN heterostructures is of great importance to lower structural uncertainties. Indeed traditional method of transferring exfoliated graphene on h-BN leads a random orientation of the latices. Although recent progresses in the field revealed that post-treatment of the samples at elevated temperatures can drive graphene to rotate and follow h-BN lattices\,\cite{Wang2016a}\cite{Woods2016}, such approaches are more efficient in sub-micrometer flakes. \textit{In-situ} growng graphene on h-BN, however, is of proven capabilities to control $\Phi$ in much larger samples.

\begin{figure}[t]
\centering
\includegraphics[width=0.4\textwidth]{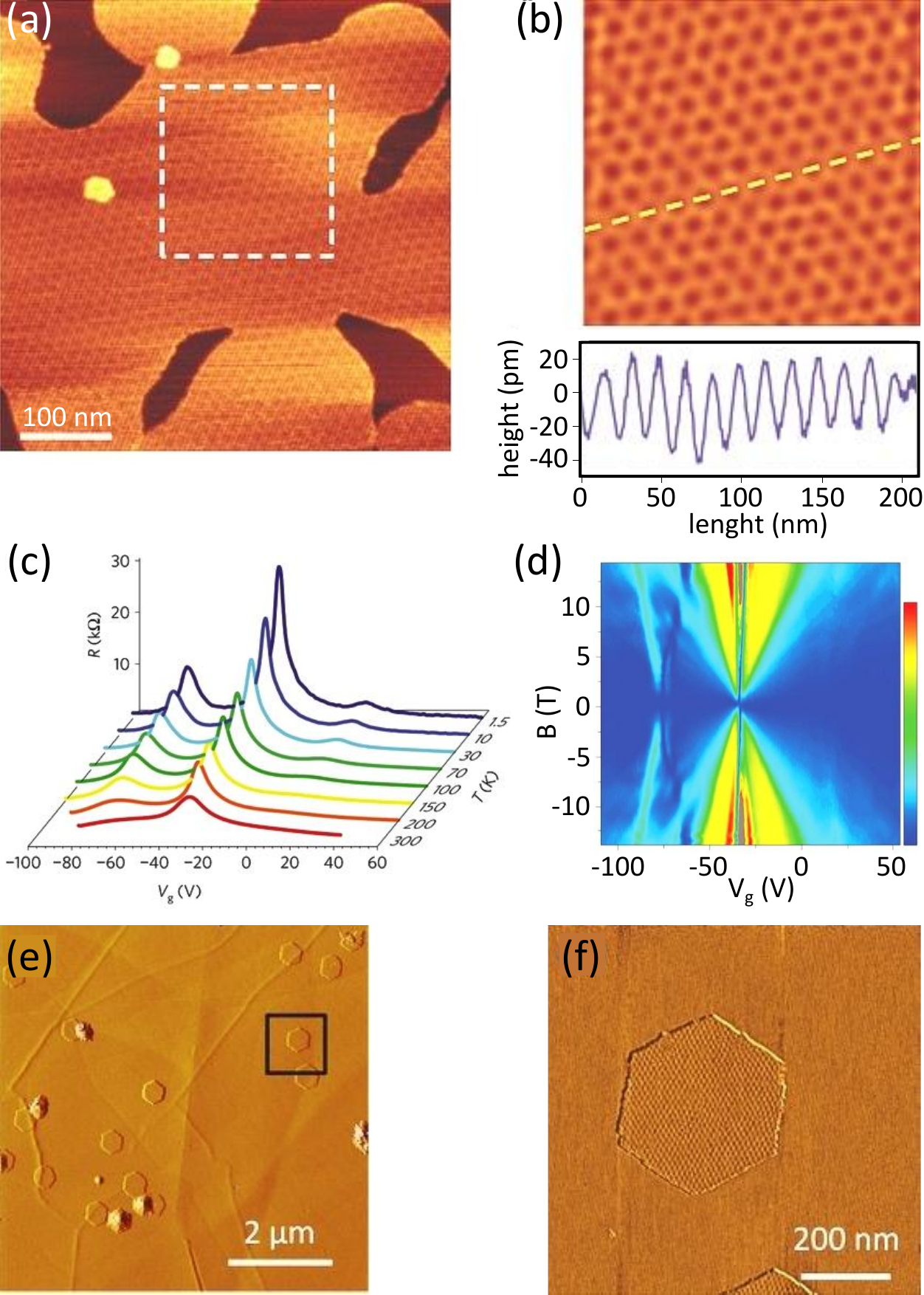} 
\centering
\caption{Crystalline biased growth of graphene on h-BN\\
a) Moir\' e pattern due to the superposition of the graphene and h-BN lattices, b) Filtered inverse fast Fourier transform of the pattern which is visible in the dashed square in a, height profile along the dashed line is shown in the lower part. The periodicity of the oscillations can be used to calculate the rotation angle between the lattices. Reprinted by permission from Macmillan Publishers Ltd: Nature Materials\,\cite{Yang2013a}, copyright (2013).\\
c) Gate dependence of the resistivity measured at different temperatures: satellite peaks shown at the left and right side of the Dirac point are due to the formation of the superlattice. d) This effect also shows up as the pattern at the left side of the fan diagram of the $R_{xy}$ in the quantum hall measurement. Reprinted by permission from Macmillan Publishers Ltd: Nature Materials\,\cite{Yang2013a}, copyright (2013).\\
e) Topography of the small graphene flakes and (f) moir\' e pattern associated to graphene/h-BN superposition. Reprinted with adaptations from\,\cite{Tang2013}.\\
The mappings shown in \emph{a}, \emph{b}, \emph{e} and \emph{f} are done with atomic force microscopy.\\
}
\label{fig: DG_literature_review_3}
\end{figure}

Yang \textit{et al} utilized plasma-enhanced CVD technique to grow graphene on mechanically exfoliated h-BN flakes\,\cite{Yang2013a}. Large area, epitaxial and single crystal graphene domains directly grown on the h-BN flakes were obtained. Breaking down the methane molecules with a remote plasma source eliminated the need for a catalyst and enhanced the growth rate and the domains size. The cleanness of the flakes was high enough that they manage to observe the moir\' e pattern associated to superposition of graphene and h-BN crystals by AFM analysis (\autoref{fig: DG_literature_review_3}-a,b). This analysis showed that graphene's lattice follows the orientation of the underlying h-BN. The size of the graphene was limited by the size of the h-BN flake, large enough to fabricate devices for transport experiments. The signature of the superposition of the lattices as extra Dirac points in the resistivity and quantum Hall effect measurements revealed at low temperatures (\autoref{fig: DG_literature_review_3}-c,d).

Note that similar alignment was reported later by Tang \textit{et al}\,\cite{Tang2013}\cite{Tang2015} (\autoref{fig: DG_literature_review_3}-e,f) and Mishra \textit{et al}\,\cite{Mishra2016}.

\begin{table*}[t]
\centering
\captionof{table}{Summary of the reports of \textit{in-situ}growing graphene on thick h-BN flakes, in chronological order}
\resizebox{.75\textwidth}{!}{%
\begin{tabular}{ccccccccc}
\hline
report & process & precursor & temperature & duration & size & thickness & orientated$^\ddagger$ & mobility \\ 
\hline 
Ding\,\cite{Ding2011}& CVD & $CH_4$ (50-90\,sccm) & 1000\degree C & 3-8\,min & not reported & > 6\,L & unclear & not reported \\

Son\,\cite{Son2011}& CVD & $CH_4$ (30-50\,sccm) & 900-1000\degree C & 2\,hrs & 100\,nm & $\approx 0.5\,nm$ & unclear & not reported\\

Tang\,\cite{Tang2012}& CVD & $CH_4$ (5\,sccm) &	1200\degree C & 1-6\,hrs & <270\,nm & ML & unclear & not reported\\ 

Garcia\,\cite{Garcia2012a} & MBE & solid carbon & 600-930\degree C & 40.6 min & nm-scale & ML & unclear & not reported \\

Yang\,\cite{Yang2013a} &	PECVD & $CH_4^\dagger$ & $\approx 500\degree C$ & $\gg$ 3\,hrs & $\mu$m-scale & ML\& BL & yes & $\sim5,000\, cm^2/Vs$ (at 1.5\,K) \\

Tang\,\cite{Tang2013} & CVD & $CH_4$ (5\,sccm) & 1200\degree C & 1-5\,hrs 	& $\mu$m-scale & ML\& BL & yes & $20,000\, cm^2/Vs$ (at 300\,K) \\

Tang\,\cite{Tang2015} & CVD & $C_2H_2^\dagger$ & 1280-1350\degree C & 5-40\,min	& $\mu$m-scale & ML & yes & $20,000\, cm^2/Vs$ (at 300\,K)\\

Mishra\,\cite{Mishra2016} & cold-wall CVD & $CH_4^\dagger$ & 1000-1150\degree C & 30\,min 	& $\mu$m-scale & ML & yes & not reported\\

Arjmandi-Tash\,\cite{Arjmandi-Tash2017} & CVD & $CH_4$ (5\,sccm) & 1050\degree C & 90\,s 	& mm-scale & ML & unclear & $20,000\, cm^2/Vs$ (at 80\,K)\\
 
Plaut\,\cite{Plaut2017} & MBE & solid carbon & 500-1000\degree C & 65\,min 	& $\mu$m-scale & ML & unclear & not reported\\

\hline
\end{tabular}
}
\vspace{1 mm}
\begin{flushleft}
\footnotesize{ML: monolayer, BL: bilayer\\$^\ddagger$ if graphene follows the orientation of underlying h-BN\\
$^\dagger$ The flow rate of the precursor was not reported in this work.}
\end{flushleft} 
\label{tbl: summary}

\end{table*}

\section{Discussion}
\autoref{tbl: summary} summarizes the important reports of \textit{in-situ} growing graphene on thick ($\gg$ monolayer) h-BN flakes in a chronological order. Chemical vapor deposition (including its derivatives) using methane precursor has been the most frequently employed method. Such reports demonstrated the growth in a wide temperature range.

The size of the resultant graphene samples and growth rate have been improved gradually over the last years. Similar improvements in controlling the thickness (number of layers) of graphene is also detectable. While the orientation of graphene in early reports were unclear, recent works reported a trend in graphene/h-BN lattice alignment. The best mobility reported for \textit{in-situ} grown samples is still much inferior than what is achieved in transfer-fabricated samples, even with CVD graphene\,\cite{Petrone2012} which points out the affect of crystalline defects. Indeed the techniques employed to compensate the lack of catalysts -- even-though successful to preserve the growth rate -- yet failed to yield crystalline qualities comparable to that of graphene grown on a catalyst.

\section{Conclusion and Perspective}
{\color{black} Outstanding electrical properties of graphene is effectively preserved from substrate-related perturbations by thick (few tens of nanometers) h-BN buffer layers. Conventional fabrication approach -- involving graphene transferring on h-BN flakes -- is prone to mechanically altering (e.g. tearing, wrinkling, stretching) the graphene. The interface contaminations by alien materials and/or air trapping are common limitations\,\cite{Uwanno2015}. The process is time-consuming and the success rate heavily depends on the expertise of the user. \textit{In-situ} growing graphene on h-BN provides a venue for achieving clean interfaces. 

Chemical vapor deposition (CVD) has been qualified as an effective \textit{in-situ} growth approach. The lack of a catalyst material -- otherwise existing in conventional CVD of graphene on a metallic foil -- is a great hindrance in the development of the process. In-fact early attempts compensated the slow chemical reaction by considerably elongating the growth up to hours and days. The new development in which the catalyst material indirectly promotes the graphene growth on h-BN\,\cite{Arjmandi-Tash2017} is a breakthrough as the full coverage of graphene on millimeter scale h-BN flakes can now be achieved in the rate identical to on catalyst. The electronic transport characteristics of graphene achieved by \textit{in-situ} growth approaches are still inferior than the devices achieved in transfer-fabrication methods. In-fact, while the charge carrier mobility of \textit{in-situ} growing graphene fails to exceed 20\,000 $cm^2/V.s$, ballistic transport with tens to hundreds times higher mobilities are now commonly achieved with transfer-fabricated graphene/h-BN heterostructures\,\cite{Mayorov2011,Wang2013a}. 
Indeed largely found crystalline defects serve as charge carrier scattering centers and strongly suppress the carrier mobility in \textit{in-situ} grown graphene. Certain measures, including the optimization of the growth parameters have to be taken to improve the crystalline order of graphene in the future. The electronic band structure of van der Waals heterostructures can be customized by engineering the twist angle between crystals (moiré superlattice potential). In-fact precise control of the twist angle achieves insulating, semi-metallic and superconducting states in such heterostructures\,\cite{Cao2018a, Codecido2019}. Here the \textit{in-situ} growth approaches offer great potentials as the orientation of growing graphene is governed by the crystalline orientation of the background h-BN. Sandwiching graphene in between two h-BN flakes provides the best device quality; no \textit{in situ} growth method has yet succeeded to yield such structures. Additionally, direct growth of van der Waals hetero-structures with multiple two-dimensional materials is on the perspective. The choice of the CVD for the \textit{in situ} growth of graphene on h-BN is largely motivated  by the proven qualification of the method in growing graphene on copper. Unlike on copper, the growth is not self-limited on h-BN and nothing guarantees a uniform monolayer film thickness. Although the observation of restricted monolayer growth on an h-BN ridge in MBE\,\cite{Plaut2017} is promising, customizing the number of layers is yet out of the reach. Besides CVD and MBE, certain efforts have to be made to evaluate the efficiency of other exiting approaches (e.g. bottom-up synthesis using polycyclic aromatic hydrocarbons) or developing novel modalities in \textit{in situ }growing graphene on h-BN in the future.
}



\balance


\bibliography{library} 
\bibliographystyle{rsc} 

\end{document}